\documentclass{iopart}
\usepackage{iopams}

\begin{document}

\title{Spin-orbit coupling and the conservation of angular momentum}
\author{V Hnizdo}

\address{National Institute for Occupational Safety and Health,\\
Morgantown, West Virginia 26505, USA}
\ead{vhnizdo@cdc.gov}
\begin{abstract}
In nonrelativistic quantum mechanics, the total (i.e.\ orbital plus spin)     angular momentum of a charged particle with spin that moves in a Coulomb plus spin-orbit-coupling  potential is conserved. In a classical nonrelativistic   treatment of this problem, in which the Lagrange equations determine the      orbital motion and the Thomas equation yields the rate of change of the spin, the particle's total angular momentum in which the orbital angular momentum is defined in terms of the kinetic momentum is generally not conserved. However, a generalized total angular momentum, in which the orbital part is defined in terms of the canonical momentum, is conserved. This illustrates the fact that the quantum-mechanical operator of momentum corresponds to the canonical      momentum of classical mechanics.                                              
\end{abstract}

\section{Introduction}
Quantum-mechanical analysis of atomic spectra is based on the conservation of the total angular momentum, defined as the sum of the orbital angular momentum and intrinsic spin,
of an electron moving in a central plus the associated spin-orbit-coupling potential (see, e.g., \cite{LLQM}). This conservation  follows immediately from the fact that, in nonrelativistic quantum mechanics, the operators of the Cartesian components and the square of the total angular momentum commute with the operator of the spin-orbit coupling and, thus, with the Hamiltonian of the system.
However, the orbital angular momentum and the spin themselves are not conserved separately since the operators of their components do not commute with the spin-orbit-coupling operator (though  the operators of their squares do).

In this paper, we consider a classical nonrelativistic treatment of this problem using a Lagrangian in which the spin and orbital degrees of freedom of a charged particle with spin moving in a central electric field are coupled through a spin-orbit interaction  $V_{\rm so}(r){\bi s} \bdot {\bi l}$, where $\bi s$  and ${\bi l}={\bi r}\times m {\bi v}$ are the particle's spin and orbital angular momenta. The resulting Lagrange equations  determine both the orbital motion and the motion of the spin. The equation of spin motion obtained is the nonrelativistic limit of the Thomas equation, derived by Thomas in 1927  in his historic, still essentially classical, analysis of the motion of a spinning electron in a Coulomb field \cite{Thomas2}. In light of the quantum-mechanical result, it would not be unreasonable to expect that this classical treatment will yield the conservation of the total angular momentum ${\bi l} +{\bi s}$.

It turns out, however, that, in this problem,  the angular momentum ${\bi l} +{\bi s}$ is in general not conserved. But we find that the total angular momentum is conserved in general when its orbital part is defined as ${\bi L}= {\bi r}\times{\bi P}$,
where $\bi P$ is the canonical (generalized) momentum. When it is realized that  canonical momentum is the classical counterpart of the quantum-mechanical operator of momentum, the classical and quantum-mechanical treatments are in harmony  after all.

The role of canonical momentum in the conservation laws of classical mechanics has been  highlighted in a pedagogical paper earlier \cite{Frank-Trigg}, but a comparison with quantum mechanics does not seem to have been used in this context yet.
Students typically encounter spin-orbit interaction only in courses on atomic physics and quantum mechanics, and thus an impression is easily created that such an interaction is a purely quantum phenomenon. A classical exposition of the spin-orbit interaction and its
effect on the conservation of angular momentum should therefore be instructive. The present topic may serve
as an illustration of the relationship between the results of classical and quantum-mechanical treatments of problems in an undergraduate course on atomic physics, assuming only some knowledge of the basic Lagrangian  formalism.

\section{Classical equations of motion of a charged particle  with spin in a Coulomb field}
Classical relativistic theory of combined spin and orbital motion (see, e.g., \cite{Corben,Barut,SS,Grassb,TB,Skag}) is a difficult subject in which one has to deal with complications such as that the centres of mass and charge of a relativistic spinning point particle do not coincide, which can lead to the so-called zitterbewegung (`jittery', or `quivery' motion), identified in force-free solutions of the Dirac equation  by Schr\"{o}dinger  \cite{zbw}, but observed experimentally only very recently, in a quantum simulation of the Dirac equation \cite{Nature2010}.

However,  a simple non-relativistic limit of a fully relativistic Lagrangian for a classical charged particle with spin can be derived.  In this limit, the  spin is parameterized in terms of vectors $\bi a$ and $\bi b$ as ${\bi s}={\bi a}\times{\bi b}$, and the following Lagrangian is obtained for a classical  particle of mass $m$, charge $q$ and nonzero spin, moving
with a velocity $\bi v$ in  the electromagnetic field ${\bi E},{\bi B}$ with potentials $\phi,{\bi A}$  \cite{Grassb}:
\begin{equation}
\fl
{\cal L}=\frac{1}{2}\,mv^2 -q\phi +\frac{q}{c}\,{\bi v}{\bdot} {\bi A}
+\frac{1}{2}\,({\bi b}\bdot \dot{\bi a}-\dot{\bi b}\bdot{\bi a})
+({\bi a}{\times} {\bi b})
\bdot\!\left[\frac{g q}{2mc}\left({\bi B}-\frac{\bi v}{c}{\times}{\bi E}\right)-
\frac{\dot{\bi v}{\times}{\bi v}}{2c^2}\right]\!.
\label{GrassL}
\end{equation}
Here, the dots on the symbols denote time derivatives and $g$ is the so-called gyromagnetic factor ($g$-factor for short), in terms of which the magnetic dipole moment $\bmu$ of the particle is expressed as
\begin{equation}
\bmu=\frac{g q}{2mc}\,{\bi s}.
\label{mu}
\end{equation}
While both the magnitude of the spin $\bi s$  and the $g$-factor are arbitrary, these quantities remain constant for a given classical particle; unlike in quantum mechanics, all the Cartesian components of the vectors $\bi s$ and $\bmu$ are well defined.
In order that Lagrangian (\ref{GrassL}) can be used for setting up equations of motion, the acceleration $\dot{\bi v}$ in its spin-dependent part is understood to be that which would arise in the absence of a spin-dependent interaction, i.e.\ $\dot{\bi v}=(q/m)({\bi E}+{\bi v}\times{\bi B}/c)$ (see the discussion leading to equations (11.163) and (11.168) in \cite{Jack}).

Applying now Lagrangian (\ref{GrassL}) to a particle of charge  $q=-e$
($e>0$), moving in a Coulomb potential  $\phi=e/r$, so that
${\bi A}=0$, ${\bi E}= -\bnabla(e/r)=e{\bi r}/r^3$ and ${\bi B}=0$, and
taking accordingly $\dot{\bi v}$ in the spin-dependent part of the interaction as $\dot{\bi v}= -e{\bi E}/m= -e^2{\bi r}/mr^3$ gives
\begin{eqnarray}
\fl{\cal L}&=\frac{1}{2}\,mv^2 +\frac{e^2}{r}
+\frac{1}{2}\,({\bi b}\bdot\dot{\bi a}-\dot{\bi b}\bdot{\bi a})-
\frac{(g-1) e^2}{2mc^2r^3}\,({\bi a}\times{\bi b})\bdot({\bi r}\times {\bi v})\nonumber \\
\fl&=\frac{1}{2}\,mv^2 +\frac{e^2}{r}
+\frac{1}{2}\,({\bi b}\bdot \dot{\bi a}-\dot{\bi b}\bdot{\bi a})-
V_{\rm so}(r) {\bi s}\bdot{\bi l}.
\label{L}
\end{eqnarray}
Here, ${\bi l}= {\bi r}\times m{\bi v}$ is the particle's orbital angular momentum defined in terms of the kinetic momentum $m{\bi v}$, and $V_{\rm so}(r)$, the strength  of the spin-orbit potential, is given by
\begin{equation}
V_{\rm so}(r)=\frac{(g-1) e^2}{2m^2c^2r^3}.
\label{VDirac}
\end{equation}
For $g=2$, which is to a very good approximation the $g$-factor of the electron, the spin-orbit strength (\ref{VDirac}) has the empirically correct value that accounts for the fine structure of the spectrum of the hydrogen atom (see, e.g., \cite{LLQED}, section 34).

Before we shall proceed with deriving the equations of motion from Lagrangian
(\ref{L}), a digression concerning the spin-orbit-interaction strength (\ref{VDirac}) is in order.
According to classical electrodynamics, the coupling of the spin and orbital degrees of freedom arises through the interaction
\begin{equation}
U_d= -{\bi d}\bdot{\bi E}
\label{Up}
\end{equation}
of an electric field ${\bi E}$ with the electric dipole moment
\begin{equation}
{\bi d}=\frac{1}{c}\,({\bi v}\times\bmu)
\label{p}
\end{equation}
that is relativistically acquired by a moving magnetic dipole $\bmu$ \cite{PP,Fisher}\footnote{Interestingly, the
Schwinger scattering \cite{Schwing}, which is the scattering of neutrons by the electric field of an atomic nucleus, also arises through  electric dipole moment (\ref{p}) via interaction (\ref{Up}).
The Schwinger-scattering  Hamiltonian (see \cite{LLQED}, equation  (42.1)) is transcribed classically in terms of the canonical momentum $\bi P$ as
$H=P^2/2m -{\bi P}\bdot(\bmu\times{\bi E})/mc$, which can be shown easily to be derivable from a  Lagrangian $L= m v^2/2 +({\bi v}\times\bmu)\bdot{\bi E}/c$
\cite{comment_Khol}.}.
Putting ${\bi E}=e{\bi r}/r^3$ and using (\ref{mu}) to express the magnetic moment $\bmu$ in terms of the spin $\bi s$, interaction (\ref{Up}) is evaluated as
\begin{eqnarray}
U_d&=-\frac{1}{c}\,({\bi v}\times\bmu)\bdot\frac{e\bi r}{r^3}\nonumber\\
&=-\frac{e}{c r^3}\,\bmu\bdot({\bi r}\times{\bi v})\nonumber\\
&=\frac{ge^2}{2m^2c^2r^3}\,{\bi s}\bdot{\bi l}.
\label{Upp}
\end{eqnarray}
However, for $g=2$, the strength of this spin-orbit interaction is by a factor of 2 greater than the empirically correct value (\ref{VDirac}).

This discrepancy was removed in the early days of quantum mechanics by the famous Thomas precession factor of $\frac{1}{2}$
\cite{Thomas2,Thomas1}, which is an effect of the relativistic kinematics of curvilinear motion. Due to the noncommutativity of Lorentz transformations, two successive
Lorentz boosts are in general equivalent to a single one plus a three-dimensional rotation. As a result, the rest frame of a particle moving along a curved path rotates with respect to the laboratory frame with an angular frequency that is given to first order in $v/c$ as
\begin{equation}
\bomega_{\rm T}=\frac{\dot{\bi v}\times{\bi v}}{2c^2}.
\label{omegaT}
\end{equation}
(An instructive exposition of the Thomas precession can be found in \cite{Jack}, section 11.8.) In Lagrangian (\ref{GrassL}), the Thomas precession is accounted for by the interaction term
$-({\bi a}\times {\bi b})\bdot(\dot{\bi v}\times{\bi v})/2c^2=-{\bi s}\bdot\bomega_{\rm T}$. This gives rise to the  term $-1$ in the factor $g-1$ of the empirically correct spin-orbit strength (\ref{VDirac}).
As we shall see, all spin-dependent  quantities will carry a factor $(g-1)$ instead of just $g$ when the effect of the Thomas precession is included.

We now turn to the particle's dynamics. With Lagrangian (\ref{L}), we have
\begin{eqnarray}
\frac{\partial{\cal L}}{\partial v_i}=mv_i+ mV_{\rm so}(r)({\bi r}\times {\bi s})_i,
\label{dLdv}\\
\frac{\partial{\cal L}}{\partial x_i}=-\frac{e^2}{r^3}\,x_i -m\frac{{\rm d}V_{\rm so}(r)}{dr}\frac{x_i}{r}\,{\bi r}\bdot
({\bi v}\times {\bi s})- mV_{\rm so}(r)({\bi v}\times {\bi s})_i.
\end{eqnarray}
Here, the evaluation of the partial derivatives was facilitated by the identities
\begin{equation}
{\bi s}\bdot({\bi r}\times {\bi v})={\bi v}\bdot({\bi s}\times {\bi r})
={\bi r}\bdot({\bi v}\times {\bi s}).
\end{equation}
The Lagrange equations
\begin{equation}
\frac{\rm d}{{\rm d}t}\frac{\partial{\cal L}}{\partial v_i}= \frac{\partial{\cal L}}{\partial x_i}, \quad\quad i=1,2,3
\end{equation}
then give the following force on the particle, understood as mass times acceleration:
\begin{equation}
\fl m\dot{\bi v}=-\frac{e^2}{r^2}\,\hat{\bi r}
-m\frac{{\rm d}V_{\rm so}(r)}{dr}\,[{\bi r}\bdot
({\bi v}\times {\bi s})]\hat{\bi r}- mV_{\rm so}(r)({\bi v}\times {\bi s})
+\frac{\rm d}{{\rm d}t}[mV_{\rm so}(r)({\bi s} \times {\bi r})].
\label{force}
\end{equation}

It is noteworthy that except for the Thomas-precession term $-1$ in the factor $(g-1)$ of the spin-orbit strength $V_{\rm so}(r)$,
the spin-dependent part of force (\ref{force})  can be obtained by transforming the force on the particle's magnetic dipole from its instantaneous rest frame to the laboratory frame.
Denoting instantaneous-rest-frame quantities by primes, the force on a magnetic dipole is given by \cite{Vaidman,VH92}
\begin{eqnarray}
{\bi F}' &=&\bnabla'(\bmu'\bdot{\bi B}')-
\frac{\rm d}{{\rm d} t'}\left(\frac{1}{c}\,\bmu'\times {\bi E}'\right)\nonumber\\
&=&(\bmu'\bdot\bnabla'){\bi B}'-\frac{1}{c}\,
\frac{{\rm d}\bmu'}{{\rm d}t'}\times{\bi E}' +\frac{4\pi}{c}\,\bmu'\times{\bi J}',
\label{F'}
\end{eqnarray}
where ${\bi B}'$ is the magnetic field and $(\bmu'\times {\bi E}')/c$ is the so-called hidden momentum of the dipole (see, e.g., \cite{VH97} and references therein);
the second line gives an equivalent expression obtained using a vector identity and the Amp\`ere-Maxwell law\footnote{Dropping the primes, $\bnabla(\bmu\bdot{\bi B})=(\bmu\bdot\bnabla){\bi B}+\bmu\times(\bnabla\times{\bi B})$ ($\bmu$ is not a function of spatial coordinates); $\bnabla\times{\bi B}=4\pi{\bi J}/c+\partial{\bi E}/c\partial t$.}. In our case, the external electric current density $\bi J'$ vanishes at the dipole's location and, to first order in $v/c$,
${\bi B}'= -({\bi v}\times{\bi E})/c$ and
${\bi E}'={\bi E}$, since the laboratory-frame magnetic field $\bi B$ is
zero\footnote{See \cite{Jack}, equation (11.149).};
also, to first order in $v/c$, $v(\bmu'\bdot\bnabla')=v(\bmu
\bdot\bnabla)$ and ${\rm d}\bmu'/{\rm d}t'
={\rm d}\bmu/{\rm d}t$\footnote{This can be shown using equations (11.22) and (11.76) of \cite{Jack} and equation (7) of \cite{Fisher}. (Note the different use of primes in the latter reference.)}.
Equation (\ref{F'}) thus  yields
\begin{equation}
{\bi F}'={\bi F}+O(v^2/c^2),
\end{equation}
where
\begin{equation}
{\bi F} = -(\bmu\bdot
\bnabla)\left(\frac{1}{c}\,{\bf v}
\times{\bi E}\right)+\frac{1}{c}\,{\bi E}\times\frac{{\rm d}\bmu}{{\rm d}t}.
\label{F}
\end{equation}
For the Coulomb field ${\bi E}=e{\bi r}/r^3$, the force (\ref{F}) can now be shown by
straightforward vector algebra (see appendix A) to equal the spin-dependent part of the force (\ref{force}) with $V_{\rm so}(r)=ge^2/(2m^2c^2r^3)$ and the spin $\bi s$
expressed in terms of the magnetic moment $\bmu$  using equation (\ref{mu}). We note that the force on a moving
magnetic dipole derived in \cite{AlJaber} is given by the same expression as  that of equation (\ref{F}).

The partial derivatives of Lagrangian (\ref{L}) with respect to the spin variables and their time derivatives  are
\begin{equation}
\frac{\partial \cal{L}}{\partial {\bi a}}
=-\frac{1}{2}\,\dot{\bi b}- V_{\rm so}(r)\,{\bi b}\times {\bi l},\quad\;\;
\frac{\partial \cal{L}}{\partial \dot{\bi a}}=\frac{1}{2}\,{\bi b}
\label{dL/da}
\end{equation}
and
\begin{equation}
\frac{\partial \cal{L}}{\partial{\bi b}}=\frac{1}{2}\,\dot{\bi a} +V_{\rm so}(r)\,{\bi a}\times {\bi l},\quad\;\;
\frac{\partial \cal{L}}{\partial\dot{\bi b}}=-\frac{1}{2}\,{\bi a}.
\label{dL/db}
\end{equation}
The Lagrange equations for the spin degrees of freedom
\begin{equation}
\frac{\rm d}{{\rm d}t}\frac{\partial\cal{L}}{\partial \dot{\bi a}}= \frac{\partial\cal{L}}{\partial {\bi a}},\quad\;\; \frac{\rm d}{{\rm d}t}\frac{\partial\cal{L}}{\partial \dot{\bi b}}= \frac{\partial\cal{L}}{\partial {\bi b}}
\end{equation}
therefore yield
\begin{equation}
\dot{\bi b}= V_{\rm so}(r)\,{\bi l}\times {\bi b},\quad\;\;
\dot{\bi a}= V_{\rm so}(r)\,{\bi l}\times {\bi a},
\end{equation}
and thus, since $\dot{\bi s}= \dot{\bi a}\times{\bi b}+{\bi a}\times\dot{\bi b}$,
\begin{eqnarray}
\dot{\bi s}&=V_{\rm so}(r)\,[({\bi l}\times{\bi a})\times{\bi b}+
{\bi a}\times({\bi l}\times{\bi b})]\nonumber\\
&=V_{\rm so}(r)\,{\bi l}\times({\bi a}\times{\bi b})\nonumber\\
&=mV_{\rm so}(r)\,({\bi r}\times {\bi v})\times{\bi s}.
\label{sdot}
\end{eqnarray}

The rate of change of spin (\ref{sdot}) is that of the nonrelativistic limit of the Thomas equation \cite{Thomas2}, which  is a form of  the well-known relativistic Bargmann--Michel--Telgedi (BMT)                                                               equation  for an arbitrary classical spin and $g$-factor \cite{BMT} (instructive derivations of the BMT and Thomas equations are given in \cite{Jack}, section 11.11). The Thomas equation to first order in $v/c$ reads \begin{equation}
\dot{\bi s} =\frac{gq}{2mc}\,{\bi s}\times\left({\bi B}-\frac{1}{c}\,
{\bi v}\times{\bi E}\right)+\frac{q}{mc}\,{\bi s}\times\frac{{\bi v}\times{\bi E}}{2c}.
\label{Thomas}
\end{equation}
In our case, $q=-e$, ${\bi E}=e{\bi r}/r^3$ and ${\bi B}=0$. With these values, equation (\ref{Thomas}) simplifies to
\begin{equation}
\dot{\bi s}=\frac{(g-1)e^2}{2mc^2r^3}({\bi r}\times{\bi v})\times{\bi s},
\label{sdotT}
\end{equation}
which is indeed the same as equation (\ref{sdot}), recalling that
the spin-orbit interaction strength $V_{\rm so}(r)$ in (\ref{sdot}) is given by (\ref{VDirac}).

The physical content of the Thomas equation can be  made more transparent  by
replacing $q\bi E$ in the second term of (\ref{Thomas}) with $m\dot{\bi v}$, which is consistent with the nonrelativistic approximation employed, and then re-writing  the equation in terms of magnetic moment (\ref{mu}) and the Thomas precession frequency (\ref{omegaT}):
\begin{equation}
\dot{\bi s}=\bmu\times\left({\bi B}-\frac{1}{c}\,
{\bi v}\times{\bi E}\right)+\bomega_{\rm T}\times{\bi s}.
\label{Thomasb}
\end{equation}
We recognize in the first term on the right-hand side  the torque
$\bmu\times{\bi B}'$ on a magnetic dipole $\bmu$ by its instantaneous-rest-frame magnetic field ${\bi B}'= {\bi B}-({\bi v}\times{\bi E})/c$; the second term describes the Thomas precession of the spin.

\section{Angular momentum of a classical charged particle with spin in a Coulomb field}
The rate of change of the orbital angular momentum defined in terms of the kinetic momentum
$m{\bi v}$ as ${\bi l}= {\bi r}\times m{\bi v}$
equals the torque ${\bi r\times}m\dot{\bi v}$ due to force (\ref{force}):
\begin{eqnarray}
\fl\dot{\bi l}
&={\bi r\times}m\dot{\bi v}\nonumber\\
\fl &=-mV_{\rm so}(r){\bi r}\times({\bi v}\times {\bi s})
+{\bi r}\times\frac{\rm d}{{\rm d}t}[mV_{\rm so}(r)({\bi s} \times {\bi r})]\nonumber \\
\fl &=-2mV_{\rm so}(r){\bi r}\times({\bi v}\times {\bi s})
-m\frac{{\rm d}V_{\rm so}(r)}{{\rm d}t}\,{\bi r}\times({\bi r}\times {\bi s})
-mV_{\rm so}(r){\bi r}\times({\bi r}\times\dot{{\bi s}}).
\label{ldot}
\end{eqnarray}
Here, the fact that ${\bi r}\times \hat{\bi r}=0$ was used. Employing
explicit expression (\ref{VDirac}) for $V_{\rm so}(r)$, we have
\begin{eqnarray}
\frac{\rmd V_{\rm so}(r)}{\rmd t}&=({\bi v}\bdot\bnabla)V_{\rm so}(r)\nonumber\\
&=-\frac{3(g-1)e^2}{2m^2c^3r^5}\,{\bi v}\bdot{\bi r},
\end{eqnarray}
the use of which in (\ref{ldot})  yields
\begin{equation}
\fl\dot{\bi l}=
-\frac{(g-1)e^2}{2mc^2r^3}\,[2{\bi r}\times({\bi v}\times{\bi s})
-(3/r^2)({\bi r}\bdot{\bi v}){\bi r}\times({\bi r}\times{\bi s})
+{\bi r}\times({\bi r}\times\dot{\bi s})].
\label{lDot}
\end{equation}

Using now rates of change (\ref{sdotT}) and (\ref{lDot}),
it can be seen easily that $\dot{\bi l}+\dot{\bi s}$ vanishes, i.e.\ the total angular momentum ${\bi l}+{\bi s}$ is conserved, only in the special case of the direction of the spin $\bi s$ being normal to the orbital plane defined by the vectors $\bi r$ and $\bi v$,  and, at the same time, the orbit being circular, i.e.\ ${\bi r}\bdot{\bi v}=0$. These  are in fact the conditions for separate conservation of $\bi l$ and  $\bi s$.

According to Noether's theorem,
which associates a conserved quantity with each symmetry of the Lagrangian of a system (see, e.g., \cite{Hanc}), there is a conserved angular momentum quantity if the system's Lagrangian is  invariant  with respect to a rotation. Lagrangian (\ref{L}) is invariant under a rotation about the center of the Coulomb field; this fact can be shown to determine
the conserved angular momentum quantity as ${\bi L} + {\bi s}$, where $\bi L$ is an orbital angular momentum defined in terms of the canonical momentum ${\bi P}=\partial \cal{L}/\partial{\bi v}$ as ${\bi L}= {\bi r}\times {\bi P}$.
However, we here  prefer to demonstrate the conservation of the angular momentum
${\bi L} + {\bi s}$ by an explicit  direct calculation, leaving the application of Noether's theorem to appendix B.

Lagrangian (\ref{L}) yields a canonical momentum
\begin{eqnarray}
{\bi P}&=\frac{\partial{\cal L}}{\partial {\bi v}}
\nonumber\\
&=m {\bi v} +\frac{(g-1)e^2}{2mc^2r^3}\,({\bi r}\times {\bi s}),
\label{GrassP}
\end{eqnarray}
and the canonical orbital momentum is thus
\begin{eqnarray}
{\bi L}&= {\bi r}\times {\bi P}\nonumber \\
&={\bi r}\times m {\bi v} +\frac{(g-1)e^2}{2mc^2r^3}\,{\bi r}\times({\bi r}\times {\bi s}).
\label{canonL}
\end{eqnarray}
The rate of change of $\bi L$ is given by
\begin{eqnarray}
\fl\dot{\bi L}&=\dot{\bi l}
+\frac{\rm d}{{\rm d}t}\left[\frac{(g-1)e^2}{2mc^2r^3}\,{\bi r} \times({\bi r}\times {\bi s})\right]\nonumber\\
\fl&=\dot{\bi l}+\frac{(g{-}1)e^2}{2mc^2r^3}\,[{\bi v}{\times}({\bi r}\times{\bi s}){+}{\bi r}{\times}({\bi r}\times\dot{\bi s}){+}{\bi r}{\times}({\bi v}\times{\bi s})
{-}(3/r^2)({\bi r}{\bdot}{\bi v}){\bi r}{\times}({\bi r}\times{\bi s})],
\label{Ldot}
\end{eqnarray}
where $\dot{\bi l}$ is the rate of change (\ref{lDot}) of the kinetic angular momentum
${\bi r}\times m{\bi v}$. Using (\ref{lDot}) in (\ref{Ldot}) gives
\begin{eqnarray}
\dot{\bi L}&=-\frac{(g-1)e^2}{2mc^2r^3}\,[{\bi r}\times({\bi v}\times{\bi s})+
{\bi v}\times({\bi s}\times{\bi r})]\nonumber \\
&=-\frac{(g-1)e^2}{2mc^2r^3}\,({\bi r}\times{\bi v})\times{\bi s}.
\label{Ldot_2}
\end{eqnarray}
Combining this result with the Thomas rate of change of spin (\ref{sdotT}) yields
\begin{equation}
\dot{\bi L}+\dot{\bi s} =0.
\end{equation}
The total angular momentum ${\bi L} +{\bi s}$ is thus conserved.

\section{Concluding remarks}
It is interesting to note that Thomas deduced in his landmark paper \cite{Thomas2} that the  orbit-averaged value of the angular momentum ${\bi l}+{\bi s}$ of a spinning electron is conserved in its motion in a Coulomb field.
Since Thomas's analysis, little attention seems to have been paid to the problem  of angular momentum conservation in semiclassical models of hydrogen-like atoms. Recently, however, Lush has  revisited this problem and, using a different approach to ours, concluded that not even the orbit-averaged value of ${\bi l}+{\bi s}$ is  conserved in this problem \cite{Lush1,Lush2}. Lush attributes the reason for the different finding in \cite{Thomas2} to Thomas's omission of the contribution of hidden momentum  to the force on a magnetic dipole and suggests the Thomas precession as the source of the nonconservation.

In the problem at hand, omitting the Thomas precession term in a nonrelativistic Lagrangian  would result in the replacement of the factor $g-1$ in the spin-orbit coupling strength and the rate of change of spin by just the factor $g$. This, for the electron's $g$-factor of 2, would result in  a spin-orbit strength by a factor of 2 greater than the empirically correct one and a rate of change of spin twice as great as that of the nonrelativistic limit of the Thomas equation. Moreover, irrespective  of these flaws, the angular momentum ${\bi l}+{\bi s}$ of the orbiting electron would still not be conserved in general.

We note that  laws of
conservation of angular momentum  hold true in general only when the angular momentum of the electromagnetic field in the system concerned is included \cite{VH92a,Kiessling}.
However, it can be shown easily that in the problem considered here the angular momentum of the central Coulomb field and the magnetic field of the orbiting particle's intrinsic magnetic moment, defined with respect to the Coulomb field's centre, vanishes in the nonrelativistic limit\footnote{The angular momentum  of the fields of a magnetic dipole $\bmu$ and a charge $q$ with respect to the position of the dipole is given by
${\bi M}_{\rm f}= q{\bi r}\times(\bmu\times{\bi r})/c r^3$, where $\bi r$ is the charge's displacement from the dipole \cite{T}. But the field angular momentum
${\bi M}'_{\rm f}$ about the  position of the charge vanishes since ${\bi M}'_{\rm f}={\bi M}_{\rm f}-
{\bi r}\times {\bi P}_{\rm f}$, where ${\bi P}_{\rm f}=q(\bmu\times{\bi r})/c r^3$
is the field linear momentum \cite{T}. See also \cite{VH92a}.}. The angular momentum of the central Coulomb field and the magnetic field due to the particle's orbital magnetic moment can be shown to vanish, too \cite{Dryzek}.

The finding of our analysis is that the orbital quantity that is relevant
to  angular momentum conservation in a classical system with spin-orbit interaction
is an orbital angular momentum defined in terms of the canonical momentum, not the standard kinetic momentum. Only the sum of such orbital angular momentum  and the spin is in general conserved in the motion of a classical charged particle with spin in a Coulomb plus the associated spin-orbit potential. This result is a forceful reminder of the fact
that the quantum-mechanical operator of momentum corresponds to
the canonical momentum of classical mechanics. This fact alone ensures that there is harmony between the
quantum-mechanical and classical treatments in such an important aspect of the        problem considered here as the conservation of angular momentum.

\section*{Acknowledgments}
The author acknowledges informative correspondence with David C Lush, whose work brought this paper's problem to the author's attention.
This paper was written by the author in his private capacity. No official support or endorsement by the Centers for Disease Control and Prevention is intended or should be inferred.

\appendix
\section{}
Here, we show that force (\ref{F}) with ${\bi E}=e{\bi r}/r^3$,
\begin{eqnarray}
{\bi F} &= -(\bmu\bdot
\bnabla)\left(\frac{1}{c}\,{\bi v}
\times{\bi E}\right)+\frac{1}{c}\,{\bi E}\times\frac{{\rm d}\bmu}{{\rm d}t}\nonumber \\
&=-\frac{e}{c}\,{\bi v}\times(\bmu\bdot
\bnabla)\frac{\bi r}{r^3}
-\frac{e}{c}\,\frac{{\rm d}\bmu}{{\rm d}t}\times\frac{\bi r}{r^3},
\label{F2}
\end{eqnarray}
equals the spin-dependent part of force (\ref{force}) with $V_{\rm so}(r)=ge^2/2m^2c^2r^3$ and ${\bi s}= -(2mc/ge)\bmu$,
\begin{eqnarray}
\fl{\bi F}_s&=-m\frac{\rmd V_{\rm so}(r)}{\rmd r}[{\bi r}\bdot({\bi v}\times{\bi s})]\hat{\bi r}-mV_{\rm so}(r)
({\bi v}\times{\bi s})+\frac{\rmd}{\rmd t}[mV_{\rm so}(r)({\bi s}\times{\bi r})]\nonumber \\
\fl&=-\frac{3e}{cr^4}[{\bi r}\bdot({\bi v}\times\bmu)]\hat{\bi r}
+\frac{e}{cr^3}({\bi v}\times\bmu) -\frac{e}{c}\,\bmu\times({\bi v}\bdot\bnabla)\frac{\bi r}{r^3}
-\frac{e}{c}\,\frac{{\rm d}\bmu}{{\rm d}t}\times\frac{\bi r}{r^3}.
\label{Fs}
\end{eqnarray}

To this end, we use the differentiation formula
\begin{equation}
({\bi a}\bdot\bnabla)\frac{\bi r}{r^3}
=\frac{\bi a}{r^3}-3({\bi a}\bdot{\bi r})\frac{\bi r}{r^5}
\end{equation}
to perform the following evaluation:
\begin{eqnarray}
\fl
{\bi v}\times(\bmu\bdot\bnabla)\frac{\bi r}{r^3}
-\bmu\times({\bi v}\bdot\bnabla)\frac{\bi r}{r^3}
&=\frac{2}{r^3}({\bi v}\times\bmu)
-\frac{3}{r^5}[(\bmu\bdot{\bi r}){\bi v}-({\bi v}\bdot{\bi r})
\bmu]\times{\bi r}\nonumber\\
\fl&=\frac{2}{r^3}({\bi v}\times\bmu)-\frac{3}{r^5}[{\bi r}\times({\bi v}\times\bmu)]\times{\bi r}\nonumber\\
\fl&=-\frac{1}{r^3}({\bi v}\times\bmu)+\frac{3}{r^5}[{\bi r}\bdot({\bi v}\times\bmu)]{\bi r}.
\label{algebra}
\end{eqnarray}
Using this result in (\ref{F2}) yields (\ref{Fs}).

\section{}
Here, we show that a symmetry of Lagrangian (\ref{L}) implies the conservation of the total angular momentum defined as the sum of the canonical orbital angular momentum and spin.

The change $\delta{\cal L}$ of this Lagrangian  under the  rotation of the orbital and spin  degrees of freedom
${\bi q}_1={\bi r}$, ${\bi q}_2={\bi a}$, ${\bi q}_3={\bi b}$ and their time derivatives
about the origin ${\bi r}=0$ by an infinitesimal oriented angle $\delta\bphi$, so that $\delta{\bi q}_i=\delta\bphi\times {\bi q}_i$ and
$\delta\dot{\bi q}_i=\delta\bphi\times \dot{\bi q}_i$, $i=1,2,3$,
is given by
\begin{eqnarray}
\delta{\cal L}&=\sum_{i=1}^3\,
\left(\frac{\partial{\cal L}}{\partial{\bi q}_i}\bdot \delta{\bi q}_i
+\frac{\partial{\cal L}}{\partial\dot{\bi q}_i}\bdot \delta\dot{\bi q}_i \right)\nonumber\\
&=\sum_{i=1}^3\,[\dot{\bi p}_i\bdot (\delta\bphi\times {\bi q}_i)
+{\bi p}_i\bdot (\delta\bphi\times\dot {\bi q}_i)]\nonumber\\
&=\delta\bphi\bdot\sum_{i=1}^3\,({\bi q}_i\times\dot{\bi p}_i
+\dot{\bi q}_i\times{\bi p}_i).
\end{eqnarray}
Here, in the second line, we introduced the canonical momenta ${\bi p}_i=\partial{\cal L}/\partial\dot{\bi q}_i$ and used the Lagrange equations
$\dot{\bi p}_i
=\partial{\cal L}/\partial{\bi q}_i$, $i=1,2,3$.
Since Lagrangian (\ref{L}) is manifestly invariant under a rotation of the vectors $\bi r$, $\bi a$, $\bi b$ and their time derivatives, we must have that $\delta{\cal L}=0$, and thus
\begin{equation}
0=\sum_{i=1}^3\,({\bi q}_i\times\dot{\bi p}_i
+\dot{\bi q}_i\times{\bi p}_i)
= \frac{\rm d}{{\rm d}t}\sum_{i=1}^3\,{\bi q}_i\times{\bi p}_i
=\frac{\rm d}{{\rm d}t}\,({\bi L}+{\bi s}),
\end{equation}
where we used  the facts that ${\bi q}_1\times {\bi p}_1 ={\bi L}$,
which is the canonical orbital angular momentum (\ref{canonL}), and ${\bi q}_2\times {\bi p}_2+ {\bi q}_3\times {\bi p}_3={\bi a}\times {\bi b}={\bi s}$, which follows from (\ref{dL/da}) and (\ref{dL/db}).
This proves that the  angular momentum  ${\bi L}+{\bi s}$ is conserved.

\end{document}